\documentclass[a4paper]{report}
\usepackage[utf8]{inputenc}
\usepackage[T1]{fontenc}
\usepackage{RJournal}
\usepackage{amsmath,amssymb,array}
\usepackage{booktabs}

\usepackage{algorithm}
\usepackage{algorithmicx}
\usepackage{algpseudocode}
\DeclareMathOperator{\logit}{logit}
\usepackage{nameref}

\begin{document}

\sectionhead{Submitted to R Journal}
\volume{XX}
\volnumber{YY}
\year{20ZZ}
\month{AAAA}

\begin{article}

\title{bssm: Bayesian Inference of Non-linear and Non-Gaussian State
Space Models in R}
\author{by Jouni Helske and Matti Vihola}

\maketitle

\abstract{%
We present an R package \CRANpkg{bssm} for Bayesian
non-linear/non-Gaussian state space modelling. Unlike the existing
packages, \pkg{bssm} allows for easy-to-use approximate inference based
on Gaussian approximations such as the Laplace approximation and the
extended Kalman filter. The package accommodates also discretely observed
latent diffusion processes. The inference is based on fully
automatic, adaptive Markov chain Monte Carlo (MCMC) on the
hyperparameters, with optional importance sampling post-correction to
eliminate any approximation bias. The package implements also a direct
pseudo-marginal MCMC and a delayed acceptance pseudo-marginal MCMC using
intermediate approximations. The package 
offers an easy-to-use interface to define models with
linear-Gaussian state dynamics with non-Gaussian observation models, and
has an \CRANpkg{Rcpp} interface for specifying custom non-linear and
diffusion models.
}

\section{Introduction}\label{introduction}

State space models (SSM) are a flexible class of latent variable models
commonly used in analysing time series data \citep[cf.][]{DK2012}. There
are a number of packages available for state space modelling for
R, especially for two special cases: a
linear-Gaussian SSM (LGSSM) where both the observation and state
densities are Gaussian with linear relationships with the states, and an
SSM with discrete state space, which is sometimes called a hidden Markov
model (HMM). These classes admit analytically tractable marginal
likelihood functions and conditional state distributions (conditioned on
the observations), making inference relatively straightforward. See for
example \citep{Petris2010, Tusell2010, KFAS, seqHMM} for review of some
of the R packages dealing with these type of models. The
present R package \pkg{bssm} is designed for Bayesian
inference of general state space models with non-Gaussian and/or
non-linear observational and state equations. The package primary aim is to provide easy-to-use and fast functions for fully Bayesian inference
with common time series models such as basic structural time series model 
\citep{Harvey1989} with exogenous covariates and simple stochastic volatility models. 
The package accomodates also custom non-linear models and discretised diffusion models.

When extending the state space modelling to non-linear or non-Gaussian
models, some difficulties arise. As the likelihood is no longer
analytically tractable, computing the latent state distributions, as
well as hyperparameter estimation of the model becomes more challenging.
One general option is to use Markov chain Monte Carlo (MCMC) methods
targeting the full joint posterior of hyperparameters and the latent
states, for example by Gibbs sampling or Hamiltonian Monte Carlo.
Unfortunately, the joint posterior is typically very high dimensional
and due to the strong autocorrelation structures of the state densities,
the efficiency of such methods can be relatively poor. Another
asymptotically exact approach is based on the pseudo-marginal particle
MCMC approach \citep{andrieu-doucet-holenstein}, where the likelihood
function and the state distributions are estimated using sequential
Monte Carlo (SMC) i.e.~the particle filter (PF) algorithm. Instead of
computationally demanding Monte Carlo methods, approximation-based
methods such extended and unscented Kalman filters may be used, as well
as Laplace approximations, which are provided for example by the INLA
\citep{inla} R package. The latter are computationally appealing, but
may lead to hard-to-quantify biases of the posterior.

Some of the R packages suitable for Bayesian state space modelling
include \CRANpkg{pomp} \citep{pomp}, \CRANpkg{rbi} \citep{rbi}, \CRANpkg{nimbleSMC} \citep{nimblesmcpaper, nimblesmcpackage}, and \CRANpkg{rstan} \citep{rstan}. With the package \pkg{pomp}, user
defines the model using R or C snippets for simulation
from and evaluation of the latent state and observation level densities, allowing flexible model construction. The \pkg{rbi} package is an interface to LibBi \citep{libbi}, a standalone software with a focus on Bayesian state space modelling on high-performance computers. The \pkg{pomp} package provides several simulation-based inference
methods mainly based on iterated filtering and maximum likelihood, whereas \pkg{rbi} is typically used for Bayesian inference via particle MCMC. For a more detailed comparison of differences of \pkg{rbi}/LibBi and \pkg{pomp} with examples, see \citep{funk_king2020}. The \pkg{nimbleSMC} package contains some particle filtering algorithms which can be used in the general Nimble modelling system \citep{nimble}, whereas the \pkg{rstan} package provides an R interface to the Stan C++ package, a general statistical modelling platform \citep{Stan}. 

The key difference to the aforementioned packages and motivation behind
the present \pkg{bssm} package is to combine the use of fast
approximation-based methods with Monte Carlo correction step, leading to computationally
efficient and unbiased (approximation error free) inference of the joint posterior of hyperparameters and latent states, as suggested in
\citep{vihola-helske-franks}. In a nutshell, the method uses MCMC which
targets an approximate marginal posterior of the hyperparameters, and an
importance sampling type weighting which provides asymptotically exact
inference on the joint posterior of hyperparameters and the latent
states. In addition to this two-stage procedure, the \pkg{bssm}
supports also delayed acceptance pseudo-marginal MCMC
\citep{Christen2005} using the approximations, and direct
pseudo-marginal MCMC. To our knowledge, importance sampling and delayed
acceptance in this form are not available in other Bayesian state space
modelling packages in R.

\section{Supported models}\label{ssm}

We denote the sequence of observations \((y_1,\ldots,y_T)\) as \(y\),
and the sequence of latent state variables
\((\alpha_1,\ldots, \alpha_T)\) as \(\alpha\). The latent states
\(\alpha_t \in \mathbb{R}^d\) are typically vector-valued, whereas we
focus mainly on scalar observations \(y_t \in \mathbb{R}\)
(vector-valued observations are also supported, assuming conditional
independence (given \(\alpha_t\)) in case of non-Gaussian observations).

A general state space model consists of two parts: observation level
densities \(g_t^{(\theta)}(y_t | \alpha_t)\) and latent state transition
densities \(\mu_t^{(\theta)}(\alpha_{t+1} | \alpha_t)\). Typically both
\(g_t^{(\theta)}\) and \(\mu_t^{(\theta)}\) depend on unknown parameter
vector \(\theta\) for which we can define arbitrary prior \(p(\theta)\).

In a linear-Gaussian SSM, both \(g_t^{(\theta)}\) and
\(\mu_t^{(\theta)}\) are Gaussian densities and they depend linearly on
the current and previous state vectors, respectively. Section
\nameref{lgssm} describes a common extension to these models supported by
\pkg{bssm}, which relaxes the assumptions on observational density
\(g_t^{(\theta)}\), by allowing exponential family links, and stochastic
volatility models. While the main focus of \pkg{bssm} is in state
space models with linear-Gaussian state dynamics, there is also support
for more general non-linear models, discussed briefly in Section
\nameref{nlgssm}. Section \nameref{using-the-bssm-package} describes how arbitrary models based on these definitions are constructed in \pkg{bssm}.

\subsection{Models with linear-Gaussian state dynamics}\label{lgssm}

The primary class of models supported by \pkg{bssm} consists of SSMs
with linear-Gaussian state dynamics of form 
\begin{align*}
\alpha_{t+1} &= c_t + T_t \alpha_t + R_t \eta_t,
\end{align*} 
where \(c_t \in\mathbb{R}^d\),
\(T_t \in\mathbb{R}^{d\times d}\), and \(R_t\in\mathbb{R}^{d \times k}\)
can depend on the unknown parameters \(\theta\) and covariates. The
noise terms \(\eta_t \sim N(0, I_k)\) and \(\alpha_1 \sim N(a_1, P_1)\)
are independent. These state dynamics can be combined with the
observational level density \(g_t\) of form 
\[
g_t(y_t | d_t + Z_t \alpha_t, \phi, u_t),
\] 
where parameters \(\phi\) and the known vector \(u_t\) are
distribution specific and can be omitted in some cases. Currently,
following observational level distributions are supported:

\begin{itemize}
\item
  Gaussian distribution: \(y_t = d_t + Z_t \alpha_t + H_t \epsilon_t\)
  with \(\epsilon_t \sim N(0, I)\).
\item
  Poisson distribution:
  \(g_t(y_t | d_t + Z_t \alpha_t, u_t) = \textrm{Poisson}(u_t \exp(d_t + Z_t \alpha_t))\),
  where \(u_t\) is the known exposure at time \(t\).
\item
  Binomial distribution:
  \(g_t(y_t | d_t + Z_t \alpha_t, u_t) = \textrm{B}(u_t, \logit^{-1}(d_t + Z_t \alpha_t))\),
  where \(u_t\) is the number of trials and
  \(\logit^{-1}(d_t + Z_t \alpha_t)\) is the
  probability of the success.
\item
  Negative binomial distribution:
  \(g_t(y_t | d_t + Z_t \alpha_t, \phi, u_t) = \textrm{NB}(\exp( d_t + Z_t \alpha_t), \phi, u_t)\),
  where \(u_t \exp( d_t + Z_t \alpha_t)\) is the expected value,
  \(\phi\) is the dispersion parameter, and \(u_t\) is a known offset
  term.
\item
  Gamma distribution:
  \(g_t(y_t | d_t + Z_t \alpha_t, \phi, u_t) = \textrm{Gamma}(\exp( d_t + Z_t \alpha_t), \phi, u_t)\),
  where \(u_t \exp( d_t + Z_t \alpha_t)\) is the expected value,
  \(\phi\) is the shape parameter, and \(u_t\) is a known offset term.
\item
  Stochastic volatility model:
  \(g_t(y_t | Z_t \alpha_t) = \exp(\alpha_t / 2)\epsilon_t\), with
  \(\epsilon_t \sim N(0, 1)\). Here the state dynamics is also fixed as
  \(\alpha_{t+1} = \mu + \rho (\alpha_t - \mu) + \sigma_{\eta} \eta_t\),
  with \(\eta_t \sim N(0,1)\) and
  \(\alpha_1 \sim N(\mu, \sigma^2_{\eta} / (1-\rho^2))\).
\end{itemize}

For multivariate models, these distributions can be combined
arbitrarily, except the stochastic volatility model case which is
currently handled separately. Also for fully Gaussian model, the
observational level errors \(\epsilon_t\) can be correlated across time
series.

\subsection{Other state space models}\label{nlgssm}

The general non-linear Gaussian model in the \pkg{bssm} has following
form:

\[
\begin{aligned}
y_t &= Z(t, \alpha_t, \theta) + H(t, \alpha_t, \theta)\epsilon_t,\\
\alpha_{t+1} &= T(t, \alpha_t, \theta) + R(t, \alpha_t, \theta)\eta_t,\\
\alpha_1 &\sim N(a_1(\theta), P_1(\theta)),
\end{aligned}
\] with \(t=1,\ldots, n\), \(\epsilon_t \sim N(0,\textrm{I}_p)\), and
\(\eta \sim N(0,\textrm{I}_k)\).

The \pkg{bssm} package also supports models where the state equation
is defined as a continuous-time diffusion model of the form \[
\textrm{d} \alpha_t =
\mu(\alpha_t,\theta) \textrm{d} t +
\sigma(\alpha_t, \theta) \textrm{d} B_t, \quad t\geq0,
\] where \(B_t\) is a  Brownian motion and where \(\mu\)
and \(\sigma\) are scalar-valued functions, with the
univariate observation density \(p(y_k | \alpha_k)\) defined at integer
times \(k=1\ldots,n\).

\section{Inference methods}\label{inference-methods}

The main goal of \pkg{bssm} is to facilitate easy-to-use full
Bayesian inference of the joint posterior \(p(\alpha, \theta | y)\) for
models discussed in Section \nameref{ssm}. The inference methods implemented
in \pkg{bssm} are based on a factorised approach where the joint
posterior of hyperparameters \(\theta\) and latent states \(\alpha\) is
given as \[
p(\alpha, \theta | y) \propto p(\theta) p(\alpha, y | \theta) = p(\theta) p(y | \theta)  p( \alpha | y, \theta),
\] where \(p(y | \theta)\) is the parameter marginal likelihood and
\(p(\alpha | y, \theta)\) is the smoothing distribution.

All the inference algorithms are based on a Markov chain Monte Carlo on
the parameters \(\theta\), whose single iteration may be summarised as
follows:

\begin{algorithm}[!ht]
    \begin{algorithmic}[1]
        \State \quad Draw a proposal $\theta' \sim N(\theta^{i-1}, \Sigma_{i-1})$.
        \State \quad Calculate the (approximate) marginal likelihood $\hat{p}(y | \theta')$.
        \State \quad Accept the proposal with probability $\alpha := \min\Big\{1, \frac{p(\theta')\hat{p}(y | \theta')}{p(\theta^{i-1}) \hat{p}(y | \theta^{i-1})}\Big\}$.
        \State \quad If the proposal $\theta'$ is accepted, set $\theta^i = \theta'$. Otherwise, set $\theta^i = \theta^{i-1}$.
        \State \label{step:adapt} \quad Adapt the proposal covariance matrix $\Sigma_{i-1}\to \Sigma_i$.
    \end{algorithmic}
    \caption{One iteration of MCMC algorithm for sampling $p(\theta | y)$.}
    \label{algo:mcmc}
\end{algorithm}

The adaptation step \ref{step:adapt} in \pkg{bssm} currently
implements the robust adaptive Metropolis algorithm \citep{Vihola2012}
with fixed target acceptance rate (0.234 by default) provided by the
\CRANpkg{ramcmc} package \citep{helske-ram}. The (approximate) marginal
likelihood \(\hat{p}(y | \theta)\) takes different forms,
leading to different inference algorithms, discussed below.

\subsection{Direct inference: marginal algorithm and particle
MCMC}\label{direct-inference}

The simplest case is with a linear-Gaussian SSM, where we can use the
exact marginal likelihood \(\hat{p}(y | \theta)=p(y | \theta)\), in
which case Algorithm \ref{algo:mcmc} reduces to (an adaptive)
random-walk Metropolis algorithm targeting the posterior marginal of the
parameters \(\theta\). Inference from the full posterior may be done
using the simulation smoothing algorithm \citep{DK2002} conditional to the sampled hyperparameters.

The other `direct' option, which can be used with any model, is using
the bootstrap particle filter (BSF) \citep{gordon-salmond-smith}, which
leads to a \emph{random} \(\hat{p}(y | \theta)\) which is an unbiased
estimator of \(p(y | \theta)\). In this case, Algorithm \ref{algo:mcmc}
reduces to (an adaptive) particle marginal Metropolis-Hastings
\citep{andrieu-doucet-holenstein}. Full posterior inference is achieved
simultaneously, by picking particle trajectories based on their
ancestries as in the filter-smoother algorithm \citep{kitagawa}. Note
that with BSF, the desired acceptance rate needs to be lower, depending
on the number of particles used \citep{doucet2015}.

\subsection{Approximate inference: Laplace approximation and the
extended Kalman filter}\label{approximate-inference}

The direct BSF discussed above may be used with any non-linear and/or
non-Gaussian model, but may be slow and/or poor mixing. To alleviate
this, \pkg{bssm} provides computationally efficient (intermediate) approximate
inference in case of non-Gaussian observation models in Section
\nameref{lgssm}, and in case of non-linear dynamics in Section \nameref{nlgssm}.

With non-Gaussian models of Section \nameref{lgssm}, we use an approximating
Gaussian model \(\tilde{p}(y,\alpha | \theta)\) which is a Laplace
approximation of \(p(\alpha, y | \theta)\) following \citep{DK2000}. We
write the likelihood as follows \[
\begin{aligned}
p(y | \theta) &= \int p(\alpha, y | \theta)\textrm{d}\alpha 
= \tilde{p}(y | \theta) E\left[\frac{p(y| \alpha, \theta)}{\tilde p(y| \alpha, \theta)}\right],
\end{aligned}
\] where \(\tilde p(y | \theta)\) is the likelihood of the Laplace
approximation and the expectation is taken with respect to its
conditional \(\tilde p(\alpha|y, \theta)\) \citep{DK2012}. Indeed,
denoting \(\hat{\alpha}\) as the mode of
\(\tilde{p}(\alpha | \theta, y)\), we may write

\[
\begin{aligned}
\log p(y | \theta) 
&= \log \tilde p(y | \theta) + \log \frac{p(y| \hat \alpha, \theta)}{\tilde p(y| \hat \alpha, \theta)}+ \log E\left[\frac{p(y| \alpha, \theta) / p(y | \hat \alpha, \theta)}{\tilde p(y| \alpha, \theta) / \tilde p(y | \hat \alpha, \theta)}\right].
\end{aligned}
\] If \(\tilde{p}\) resembles \(p\) with typical values of \(\alpha\),
the latter logarithm of expectation is zero. We take
\(\hat{p}(y |\theta)\) as the expression on the right, dropping the
expectation.

When \(\hat{p}\) is approximate, the MCMC algorithm targets an
approximate posterior marginal. Approximate full inference may be done
analogously as in Section \nameref{direct-inference}, by simulating
trajectories conditional to the sampled parameter configurations
\(\theta^i\). We believe that approximate inference is often good enough
for model development, but strongly recommend using post-correction as
discussed in Section \nameref{post-correction} to check the validity of the
final inference.

In addition to these algorithms, \pkg{bssm} also supports
\(\hat{p}(y | \theta)\) based on the extended KF (EKF) or iterated EKF (IEKF)
\citep{jazwinski} which can be used for models with non-linear
dynamics (Section \nameref{nlgssm}). Approximate smoothing based on
(iterated) EKF is also supported. It is also possible to perform direct
inference as in Section \nameref{direct-inference}, but instead of the BSF,
employ particle filter based on EKF \citep{vardermerwe}.

\subsection{Post-processing by importance
weighting}\label{post-correction}

The inference methods in Section \nameref{approximate-inference} are computationally
efficient, but come with a bias. The \pkg{bssm} implements
importance-sampling type post-correction as discussed in
\citep{vihola-helske-franks}. Indeed, having MCMC samples \((\theta^i)\)
from the approximate posterior constructed as in Section
\nameref{approximate-inference}, we may produce (random) weights and latent
states, such that the weighted samples form estimators which are
consistent with respect to the true posterior \(p(\alpha,\theta|y)\).

The primary approach which we recommend for post-correction is based on
a ``\(\psi\)-APF'\,' --- a particle filter using the Gaussian
approximations of Section \nameref{approximate-inference}. In essence, this
particle filter employs the dynamics and a look-ahead strategy coming
from the approximation, which leads to low-variance estimators; see
\citep{vihola-helske-franks} and package vignettes\footnote{\url{https://cran.r-project.org/package=bssm/vignettes/psi_pf.html}}
for more detailed description. Naturally \(\psi\)-APF can also be used
in place of BSF in direct inference of Section \nameref{direct-inference}.

\subsection{Direct inference using approximation-based delayed
acceptance}\label{direct-inference-using-approximation-based-delayed-acceptance}

An alternative to approximate MCMC and post-correction, \pkg{bssm}
also supports an analogous delayed acceptance method
\citep{Christen2005, Banterle2015} (here denoted by DA-MCMC). This
algorithm is similar to \ref{algo:mcmc}, but in case of `acceptance',
leads to second-stage acceptance using the same weights as the
post-correction would; see \citep{vihola-helske-franks} for details.
Note that as in direct approach for non-Gaussian/non-linear models, the
desired acceptance rate with DA-MCMC should be lower than the default
0.234.

The DA-MCMC also leads to consistent posterior estimators, and often
outperforms the direct particle marginal Metropolis-Hastings. However,
empirical findings \citep{vihola-helske-franks} and theoretical
considerations \citep{franks-vihola} suggest that approximate inference
with post-correction may often be preferable. The \pkg{bssm} supports
parallelisation with post-correction using OpenMP, which may further promote the
latter.

\subsection{Inference with diffusion state dynamics}\label{inference-with-diffusion-state-dynamics}

For general continuous-time diffusion models, the transition densities
are intractable. The \pkg{bssm} uses Millstein time-discretisation
scheme for approximate simulation, and inference is based on the
corresponding BSF. Fine time-discretisation mesh gives less bias than
the coarser one, with increased computational complexity. The DA and IS
approaches can be used to speed up the inference by using coarse
discretisation in the first stage and then using more fine mesh in the
second stage. For comparison of DA and IS approaches in case of
geometric Brownian motion model, see \citep{vihola-helske-franks}.

\section{Using the \pkg{bssm} package}\label{using-the-bssm-package}

Main functions of \pkg{bssm} related to the MCMC sampling, approximations, and particle filtering are written in C++, with help of \pkg{Rcpp} \citep{Rcpp} and \CRANpkg{RcppArmadillo} \citep{RcppArmadillo} packages. On the R
side, the package uses S3 methods to provide a relatively unified
workflow independent of the type of the model one is working with. The
model building functions such as \code{bsm\_ng} and \code{svm} are
used to construct the model objects of same name which can be then passed to other
methods, such as \code{logLik} and \code{run\_mcmc} which compute
the log-likelihood value and run MCMC algorithm respectively. We will
now briefly describe the main functionality of \pkg{bssm}. For more
detailed descriptions of different functions and their arguments, see the corresponding documentation in R and the package vignettes.

\subsection{Constructing the model}\label{constructing-the-model}

For models with linear-Gaussian state dynamics,
\pkg{bssm} includes some predefined models such as \code{bsm\_lg}
and \code{bsm\_ng} for univariate Gaussian and non-Gaussian structural
time series models with external covariates,
for which user only needs to supply the data and priors for unknown
model parameters. In addition, \pkg{bssm} supports general model
building functions \texttt{ssm\_ulg}, \code{ssm\_mlg} for custom univariate
and multivariate Gaussian models and \code{ssm\_ung}, and
\code{ssm\_mng} for their non-Gaussian counterparts. For these models,
users need to supply their own R functions for the evaluation of the
log prior density and for updating the model matrices given the current
value of the parameter vector \(\theta\). It is also possible to avoid
defining the matrices manually by leveraging the formula interface of
the \CRANpkg{KFAS} package \citep{KFAS} together with \code{as\_bssm}
function which converts KFAS model to a \pkg{bssm} equivalent model
object. This is especially useful in case of complex multivariate models
with covariates.

As an example, consider a Gaussian local linear trend model of the form

\[
\begin{aligned}
y_t &= \mu_t + \epsilon_t,\\
\mu_{t+1} &= \mu_t + \nu_t + \eta_t,\\
\nu_{t+1} &= \nu_t + \xi_t,
\end{aligned}
\] with zero-mean Gaussian noise terms \(\epsilon_t, \eta_t, \xi_t\)
with unknown standard deviations. Using the time series of the mean annual temperature (in Fahrenheit) in New Haven, Connecticut, from 1912 to 1971 (available in the \code{datasets} package) as an example, this model can be built with \code{bsm} function as

\begin{example}
library("bssm")
data("nhtemp", package = "datasets")
prior <- halfnormal(1, 10)
bsm_model <- bsm_lg(y = nhtemp, sd_y = prior, sd_level = prior, sd_slope = prior)
\end{example}

Here we use helper function \code{halfnormal} which defines
half-Normal prior distribution for the standard deviation parameters,
with the first argument defining the initial value of the parameter, and
second defines the scale parameter of the half-Normal distribution.
Other prior options are \code{normal}, \code{tnormal} (truncated
normal), \code{gamma}, and \code{uniform}.

As an example of multivariate model, consider bivariate Poisson model
with latent random walk model, defined as

\[
\begin{aligned}
y_{i,t} &\sim \textrm{Poisson}(\exp(x_t)), \quad i = 1, 2,\\
x_{t+1} &= x_t + \eta_t,
\end{aligned}
\] with \(\eta_t \sim N(0, \sigma^2)\), and prior
\(\sigma \sim \textrm{Gamma}(2,0.01)\). This model can be built with
\code{ssm\_mng} function as

\begin{example}
# Generate observations
set.seed(1)
x <- cumsum(rnorm(50, sd = 0.2)) 
y <- cbind(                                       
  rpois(50, exp(x)), 
  rpois(50, exp(x)))

# Log prior density function
prior_fn <- function(theta) {
  dgamma(theta, 2, 0.01, log = TRUE)               
}

# Model parameters from hyperparameters
update_fn <- function(theta) {                    
  list(R = array(theta, c(1, 1, 1)))
}

# define the model
mng_model <- ssm_mng(y = y, Z = matrix(1,2,1), T = 1, 
  R = 0.1, P1 = 1, distribution = "poisson",
  init_theta = 0.1, 
  prior_fn = prior_fn, update_fn = update_fn)
\end{example}

Here the user-defined functions \code{prior\_fn} and
\code{update\_fn} define the log-prior for the model and how the model components depend on the
hyperparameters \(\theta\) respectively.

For models where the state equation is no longer linear-Gaussian, we use
pointer-based interface by defining all model components as well as
functions defining the Jacobians of \(Z(\cdot)\) and \(T(\cdot)\) needed
by the extended Kalman filter as C++ snippets. General non-linear
Gaussian model can be defined with the function \code{ssm\_nlg}.
Discretely observed diffusion models where the state process is assumed
to be continuous stochastic process can be constructed using the
\code{ssm\_sde} function, which takes pointers to C++ functions
defining the drift, diffusion, the derivative of the diffusion function,
and the log-densities of the observations and the prior. As an example
of the latter, let us consider an Ornstein--Uhlenbeck process 
\[
\textrm{d} \alpha_t = \rho (\nu - \alpha_t) \textrm{d} t + \sigma \textrm{d} B_t,
\] 
with parameters \(\theta = (\phi, \nu, \sigma) = (0.5, 2, 1)\) and
the initial condition \(\alpha_0 = 1\). For observation density, we use
Poisson distribution with parameter \(\exp(\alpha_k)\). We first
simulate a trajectory \(x_0, \ldots, x_n\) using the \code{sde.sim}
function from the \CRANpkg{sde} package \citep{sde} and use that for the
simulation of observations \(y\):

\begin{example}
library("sde")
x <- sde.sim(t0 = 0, T = 100, X0 = 1, N = 100,
  drift = expression(0.5 * (2 - x)),
  sigma = expression(1),
  sigma.x = expression(0))
y <- rpois(100, exp(x[-1]))
\end{example}

We then compile and build the model as

\begin{example}
Rcpp::sourceCpp("ssm_sde_template.cpp")
pntrs <- create_xptrs()
sde_model <- ssm_sde(y, pntrs$drift, pntrs$diffusion, 
  pntrs$ddiffusion, pntrs$obs_density, pntrs$prior, 
  c(0.5, 2, 1), 1, FALSE)
\end{example}

The templates for the C++ functions for SDE and non-linear Gaussian models can be found from the package vignettes on the CRAN\footnote{\url{https://CRAN.R-project.org/package=bssm}}.

\subsection{Markov chain Monte Carlo in \pkg{bssm}}\label{bssm-mcmc}

The main purpose of the \pkg{bssm} is to allow computationally efficient MCMC-based
inference for various state space models. For this task, a method
\code{run\_mcmc} can be used. The function takes a number of
arguments, depending on the model class, but for many of these, default
values are provided. For linear-Gaussian models, we only need to supply
the number of iterations. Using the previously created local linear trend model for the New Haven temperature data of Section \nameref{constructing-the-model}, we run an MCMC with 100,000 iterations where first 10,000 is discarded as a burn-in
(burn-in phase is also used for the adaptation of the proposal
distribution):

\begin{example}
mcmc_bsm <- run_mcmc(bsm_model, iter = 1e5, burnin = 1e4)
\end{example}

The \code{print} method for the output of the MCMC algorithms gives a
summary of the results, and detailed summaries for \(\theta\) and
\(\alpha\) can be obtained using \code{summary} function. For all MCMC
algorithms, \pkg{bssm} uses so-called jump chain representation of
the Markov chain \(X_1,\ldots,X_n\), where we only store each accepted
\(X_k\) and the number of steps we stayed on the same state. So for
example if \(X_{1:n} = (1,2,2,1,1,1)\), we present such chain as
\(\tilde X = (1,2,1)\), \(N=(1,2,3)\). This approach reduces the storage
space and makes it more computationally efficient to use importance sampling type
correction algorithms. One drawback of this approach is that the results
from the MCMC run correspond to weighted samples from the target
posterior, so some of the commonly used postprocessing tools need to be
adjusted. Of course, in case of other methods than IS-weighting, the
simplest option is to just expand the samples to typical Markov chain
using the stored counts \(N\). This can be done using the function
\code{expand\_sample} which returns an object of class \code{"mcmc"}
of the \CRANpkg{coda} package \citep{coda} (thus the plotting and
diagnostic methods of \pkg{coda} can also be used). We can also
directly transform the posterior samples to a \code{"data.frame"} object by
using \code{as.data.frame} method for the MCMC output (for
IS-weighting, the returned data frame contains additional column
\code{weights}). This is useful for example for visualization purposes
with the \CRANpkg{ggplot2} \citep{ggplot2} package:

\begin{example}
library("ggplot2")
d <- as.data.frame(mcmc_bsm, variable = "theta")
ggplot(d, aes(x = value)) + 
  geom_density(bw = 0.1, fill = "#9ebcda") + 
  facet_wrap(~ variable, scales = "free") + 
  theme_bw()
\end{example}
\begin{figure}
\includegraphics[width=\columnwidth]{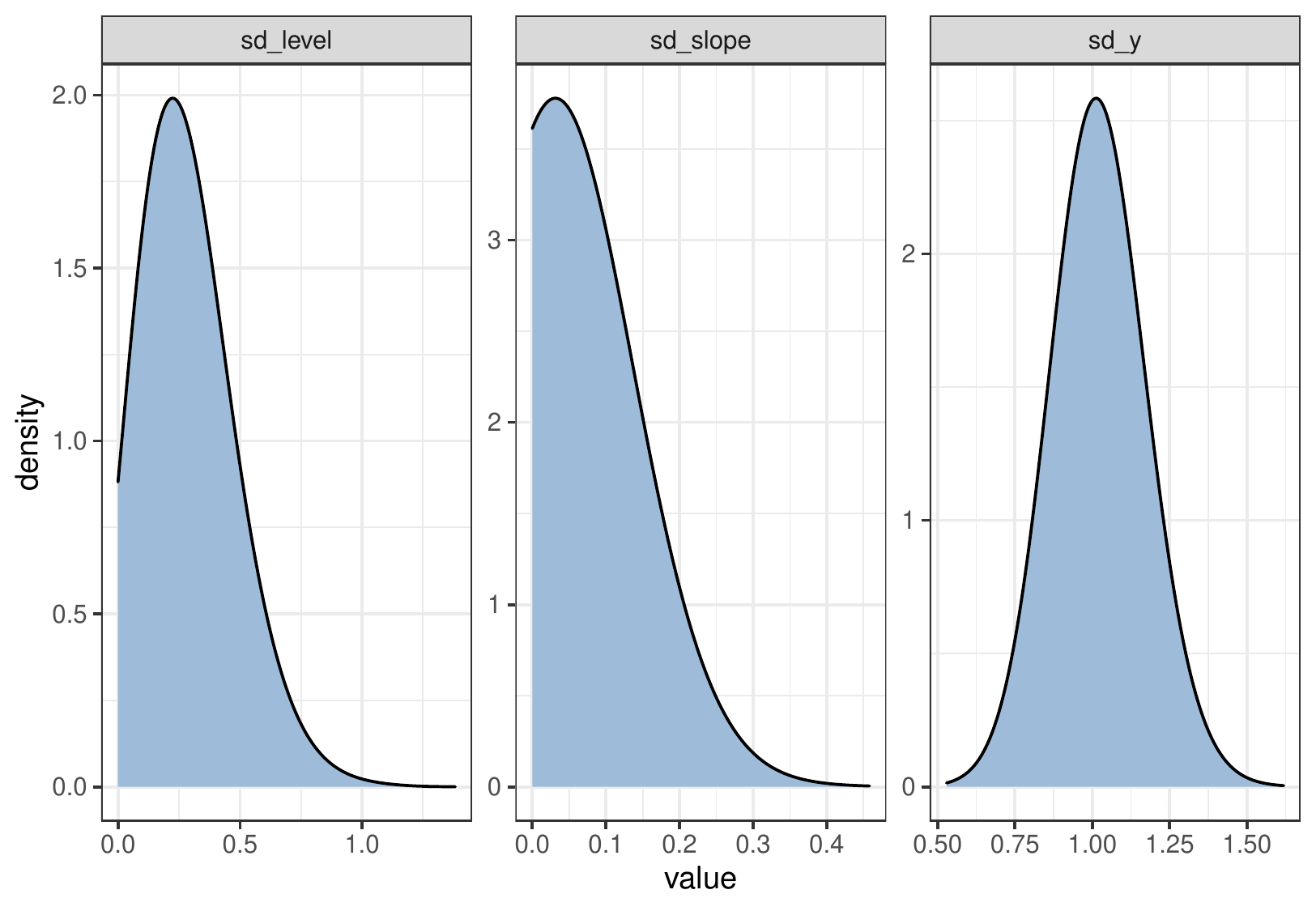} \caption[Posterior densities of linear-Gaussian model for nhtemp data]{Posterior densities of hyperparameters $\theta$ of the linear-Gaussian model for nhtemp data.}\label{fig:bsm-plot}
\end{figure}

\begin{figure}
\includegraphics[width=\columnwidth]{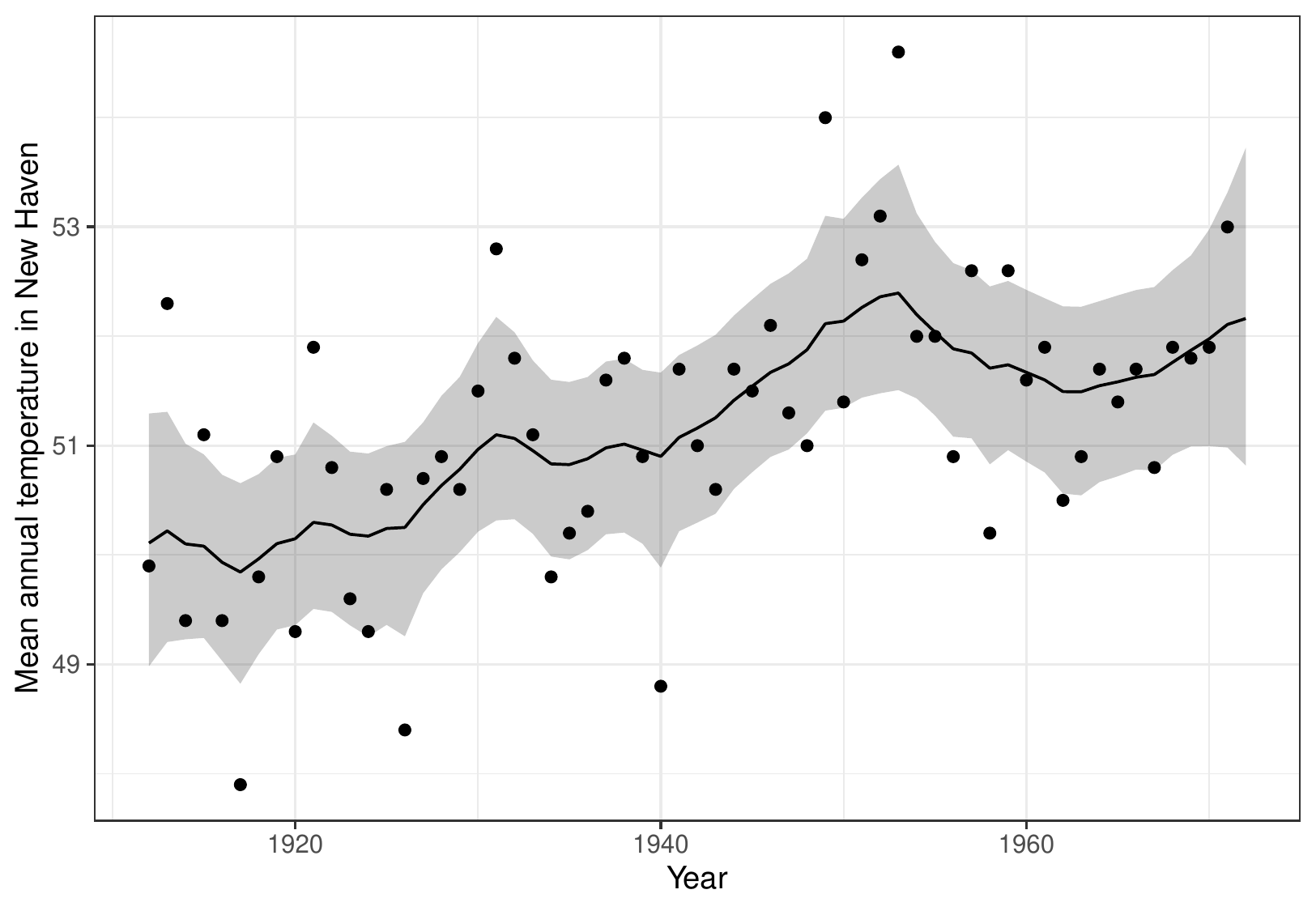} \caption[Observed annual average temperatures in New Haven (black dots) and predicted mean (solid line) with 95\% prediction intervals (grey ribbon) from `bssm`]{Observed annual average temperatures in New Haven (black dots) and predicted mean (solid line) with 95\% prediction intervals (grey ribbon) from `bssm`.}\label{fig:bsm-plot-states}
\end{figure}

Figure \ref{fig:bsm-plot} shows the estimated posterior densities of the three standard deviation parameters of the model. The relatively large observational level standard deviation $\sigma_y$ suggests that the underlying latent temperature series is much smoother than the observed series, which can also be seen from Figure \ref{fig:bsm-plot-states} which show the original observations (black dots) spread around the estimated temperature series (solid line).

\begin{example}
library("dplyr")
d <- as.data.frame(mcmc_bsm, variable = "states")
summary_y <- d 
  filter(variable == "level") 
  group_by(time) 
  summarise(mean = mean(value), 
    lwr = quantile(value, 0.025), 
    upr = quantile(value, 0.975))

ggplot(summary_y, aes(x = time, y = mean)) + 
  geom_ribbon(aes(ymin = lwr, ymax = upr), alpha = 0.25) +
  geom_line() +
  geom_point(data = data.frame(mean = nhtemp, 
    time = time(nhtemp))) +
  theme_bw() + xlab("Year") + 
  ylab("Mean annual temperature in New Haven")
\end{example}

For non-Gaussian models the default MCMC algorithm is approximate
inference based on Laplace approximation combined with importance sampling post-correction (Section
\nameref{post-correction}). It is also possible to perform first approximate
MCMC using the argument \code{mcmc\_type\ =\ "approx"}, and then
perform the post-correction step using the results from the approximate
MCMC. In doing so, we can also use the function \code{suggest\_N} to
find a suitable number of particles \(N\) for \(\psi\)-APF in the spirit
of \citet{doucet2015}:

\begin{example}
out_approx <- run_mcmc(mng_model, mcmc_type = "approx", iter = 50000)
est_N <- suggest_N(mng_model, out_approx)
out_exact <- post_correct(mng_model, out_approx, particles = est_N$N)
\end{example}

The function \code{suggest\_N} computes the standard deviation of the
logarithm of the post-correction weights (i.e.~the random part of
log-likelihood of \(\psi\)-APF) at the approximate MAP estimator of
\(\theta\) using a range of \(N\) and returns a list with component
\code{N} which is the smallest number of particles where the standard
deviation was less than one. For small and moderate problems typically
10-20 particles is enough.

\subsection{Filtering and smoothing}\label{filtering-and-smoothing}

The \pkg{bssm} also offers separate methods for performing (approximate) state filtering and smoothing which may be useful in some custom settings.

For LGSSM, methods \code{kfilter} and
\code{smoother} perform Kalman filtering and smoothing. For
non-Gaussian models with linear-Gaussian dynamics, approximate filtering
and smoothing estimates can be obtained by calls to \code{kfilter} and
\code{smoother}, in which case these functions first construct an
approximating Gaussian model for which the Kalman filter/smoother is
then applied. For non-linear models defined by \code{nlg\_ssm} we can run
approximate filtering using the extended Kalman filter with the function
\code{ekf}, the unscented Kalman filter with the function
\code{ukf}, or the iterated EKF (IEKF) by changing the argument
\code{iekf\_iter} of the \code{ekf} function. Function
\code{ekf\_smoother} can be used for smoothing based on EKF/IEKF.

For particle filtering the \pkg{bssm} package supports general
bootstrap particle filter for all
model classes of the \pkg{bssm} (function \code{bootstrap\_filter}). For \code{nlg\_ssm}, extended
Kalman particle filtering \citep{vardermerwe} is also supported (function
\code{ekpf\_filter}).

For particle smoothing, function \code{particle\_smoother} with the
smoothing based on BSF is available for all models. In addition,
\(\psi\)-APF (using argument \code{method\ =\ "psi"}) is available for
all models except of \code{ssm\_sde} class. Currently, only
filter-smoother approach \citep{kitagawa} for particle smoothing is
supported.

\section{Comparison of IS-MCMC and HMC}\label{comparison}

\citet{vihola-helske-franks} compared the computational efficiency of
delayed acceptance MCMC and importance sampling type MCMC approaches in
various settings. Here we make a small experiment comparing the generic
Hamiltonian Monte Carlo using the NUTS sampler \citep{Hoffman2014} with
\pkg{rstan}, and IS-MCMC with \pkg{bssm}. Given that the \pkg{bssm} is specialized for state space models whereas Stan is a general purpose tool suitable for wider range of problems, it is to be expected that \pkg{bssm} performs better in terms of computational efficiency. The purpose of this experiment is to illustrate this fact, i.e., that there is still demand for specialized algorithms for various types of statistical models. For
complete code of the experiment, see \nameref{appendix}.

We consider the case of a random walk with drift model with negative
binomial observations and some known covariate \(x_t\), defined as \[
\begin{aligned}
y_t &\sim \textrm{NB}(\exp(\beta x_t + \mu_t), \phi)\\
\mu_{t+1} &= \mu_t + \nu_t + \eta_t,\\
\nu_{t+1} &= \nu_t,
\end{aligned}
\] with zero-mean Gaussian noise term \(\eta_t\) with unknown standard
deviation \(\sigma_\mu\). Based on this we simulate one realization of
\(y\) and \(x\) with \(n=200\), \(\phi=5\), \(\beta= -0.9\),
\(\nu = 0.01\), \(\sigma_\mu = 0.1\).

For the IS approach we use \code{ng\_bsm} function for model
building, with prior variances 100 and 0.01 for the initial states
\(\mu_1\) and \(\nu_1\). For hyperparameters, we used fairly
uninformative half-Normal distribution with standard deviation 0.5 for
\(\sigma_\mu\) and 0.1 for \(\sigma_\nu\). We then ran the IS-MCMC
algorithm with \code{run\_mcmc} using a burn-in phase of length 10,000
and run 50,000 iterations after the burn-in, with 10 particles per SMC.

Using the same set up, we ran the MCMC with \pkg{rstan} using 15,000
iterations (with first 5000 used for warm-up). Note that in order to
avoid sampling problems, it was necessary to tweak the default control
parameters of the sampler (see Appendix).

Table \ref{tab:comparison} shows the results. We see both methods
produce identical results (within the Monte Carlo error), but while
\pkg{rstan} produces similar Monte Carlo standard errors with smaller
amount of total iterations than \pkg{bssm}, the total computation
time of \pkg{rstan} is almost 80 times higher than with \pkg{bssm} (58
minutes versus 45 seconds), which suggests that for these type of problems it is highly beneficial to take advantage of the known model structure and available approximations versus general Bayesian software such as Stan which makes no distinction between latent states $\alpha$ and hyperparameters $\theta$.

\begin{table}[h]
    \caption{Estimates of posterior mean, standard deviation and Monte Carlo standard error of the mean for hyperparameters $\theta$ and latent states for last time point for the example model.}
    \label{tab:comparison}
    \centering
    \begin{tabular}{l|rrr|rrr}
    	\toprule
        &       \multicolumn{3}{c|}{\pkg{bssm}}  &   \multicolumn{3}{c}{\pkg{rstan}} \\
        \midrule
          & Mean & SD & MCSE & Mean & SD & MCSE \\
        \midrule
        $\sigma_{\mu}$ & $0.092$  & $0.037$ & $9\times 10^{-4}$ & $0.090$  & $0.036$ & $9\times 10^{-4}$\\
        $\sigma_{\nu}$ & $0.003$  & $0.003$ & $5\times 10^{-5}$ & $0.003$  & $0.003$ & $7\times 10^{-5}$ \\
        $\phi$         & $5.392$  & $0.910$ & $2\times 10^{-2}$ & $5.386$  & $0.898$ & $1\times 10^{-2}$\\
        $\beta$        & $-0.912$ & $0.056$ & $1\times 10^{-3}$ & $-0.911$ & $0.056$ & $7\times 10^{-4}$ \\
        $\mu_{200}$    & $6.962$ & $0.346$ & $5\times 10^{-3}$ & $6.965$ & $0.349$ & $4\times 10^{-3}$ \\
        $\nu_{200}$    & $0.006$ & $0.020$ & $3\times 10^{-4}$ & $0.006$ & $0.019$ & $2\times 10^{-4}$ \\
\bottomrule    
\end{tabular}
\end{table}

\section{Conclusions}\label{conclusions}

State space models are a flexible tool for analysing a variety of time
series data. Here we introduced the R package \pkg{bssm} for
fully Bayesian state space modelling for a large class of models with
several alternative MCMC sampling strategies. All computationally
intensive parts of the package are implemented with C++ with
parallel computation support for IS-MCMC making it an attractive option
for many common models where relatively accurate Gaussian approximations
are available.

Compared to early versions of the \pkg{bssm} package, the option to
define R functions for model updating and prior evaluation have
lowered the bar for analysing custom models. The package is also written
in a way that it is relatively easy to extend to new model types similar
to current \code{bsm\_lg} in future. The \pkg{bssm} package could
be expanded to allow other proposal adaptation schemes such as adaptive
Metropolis algorithm by \citet{haario}, as well as support for
multivariate SDE models and automatic differentiation for EKF-type
algorithms.

\section{Acknowledgements}\label{acknowledgements}

This work has been supported by the Academy of Finland research grants
284513, 312605, 315619, 311877, and 331817.

\bibliography{helske-vihola}

\appendix

\section{Appendix: Code for section \nameref{comparison}}\label{appendix}

\begin{example}
library("bssm")

# Simulate the data
set.seed(123)
n <- 200
sd_level <- 0.1
drift <- 0.01
beta <- -0.9
phi <- 5

level <- cumsum(c(5, drift + rnorm(n - 1, sd = sd_level)))
x <- 3 + (1:n) * drift + sin(1:n + runif(n, -1, 1))
y <- rnbinom(n, size = phi, mu = exp(beta * x + level))

# Construct model for bssm
bssm_model <- bsm_ng(y, 
  xreg = x,
  beta = normal(0, 0, 10),
  phi = halfnormal(1, 10),
  sd_level = halfnormal(0.1, 1), 
  sd_slope = halfnormal(0.01, 0.1),
  a1 = c(0, 0), P1 = diag(c(10, 0.1)^2), 
  distribution = "negative binomial")

# run the MCMC
fit_bssm <- run_mcmc(bssm_model, iter = 60000, burnin = 10000,
  particles = 10, seed = 1)

# create the Stan model

library("rstan")

stan_model <- "
data {
  int<lower=0> n;             // number of data points
  int<lower=0> y[n];          // time series
  vector[n] x;                // covariate
}

parameters {
  real<lower=0> sd_slope;
  real<lower=0> sd_level;
  real beta;
  real<lower=0> phi;
  // instead of working directly with true state variables
  // it is often suggested use standard normal variables in sampling
  // and reconstruct the true parameters in transformed parameters block
  // this should make sampling more efficient although coding the model 
  // is less intuitive.
  vector[n] level_std;        // N(0, 1) level noise
  vector[n] slope_std;        // N(0, 1) slope noise
}
transformed parameters {
  vector[n] level;
  vector[n] slope;
  // construct the actual states
  level[1] = 10 * level_std[1];
  slope[1] = 0.1 * slope_std[1];
  slope[2:n] = slope[1] + cumulative_sum(sd_slope * slope_std[2:n]);
  level[2:n] = level[1] + cumulative_sum(slope[1:(n-1)]) + 
               cumulative_sum(sd_level * level_std[2:n]);
}
model {
  beta ~ normal(0, 10);
  phi ~ normal(0, 10);
  sd_slope ~ normal(0, 0.1);
  sd_level ~ std_normal();
  // standardised noise terms
  level_std ~ std_normal();
  slope_std ~ std_normal();
  
  y ~ neg_binomial_2_log(level + beta * x, phi);
}

"

stan_data <- list(n = n, y = y, x = x)

stan_inits <- list(list(sd_level = 0.1, sd_slope = 0.01, phi = 1, beta = 0))

# need to increase adapt_delta and max_treedepth in order to avoid divergences
fit_stan <- stan(model_code = stan_model,
  data = stan_data, iter = 15000, warmup = 5000,
  control = list(adapt_delta = 0.99, max_treedepth = 12),
  init = stan_inits, chains = 1, refresh = 0, seed = 1)

d_stan <- summary(fit_stan, pars = 
    c("sd_level", 
      "sd_slope", 
      "phi", 
      "beta",
      "level[200]",
      "slope[200]"
    ))$summary[,c("mean", "sd", "se_mean")]

d_bssm <- summary(fit_bssm, variable = "both", return_se = TRUE)

# Parameter estimates:
d_stan
d_bssm$theta
d_bssm$states$Mean[200,]
d_bssm$states$SD[200,]
d_bssm$states$SE[200,]

# Timings:
sum(get_elapsed_time(fit_stan))
fit_bssm$time[3]
\end{example}

\address{%
Jouni Helske\\
Department of Mathematics and Statistics\\
University of Jyväskylä\\ 
Finland\\
ORCiD: 0000-0001-7130-793X\\
\email{jouni.helske@jyu.fi}
}

\address{%
Matti Vihola\\
Department of Mathematics and Statistics\\
University of Jyväskylä\\ 
Finland\\
ORCiD: 0000-0002-8041-7222\\
\email{matti.s.vihola@jyu.fi}
}

\end{article}

\end{document}